\documentclass[english,twocolumn,superscriptaddress,floatfix,
  longbibliography]{revtex4-1}

\usepackage{soul}
\usepackage[dvipsnames]{xcolor}

\usepackage{xcolor}
\usepackage{graphicx}
\usepackage{amsmath}
\usepackage{amsfonts}
\usepackage{color}
\usepackage{babel}
\usepackage{url}
\usepackage{hyperref}
\usepackage{tikz}
\usepackage{CJKutf8}
\usepackage[title]{appendix}
\usepackage{siunitx}

\definecolor{lime}{HTML}{A6CE39}
\DeclareRobustCommand{\orcidicon}{
	\begin{tikzpicture}
	\draw[lime, fill=lime] (0,0) 
	circle [radius=0.16] 
	node[white] {{\fontfamily{qag}\selectfont \tiny ID}};
	\draw[white, fill=white] (-0.0625,0.095) 
	circle [radius=0.007];
	\end{tikzpicture}
	\hspace{-2mm}
}

\foreach \x in {A, ..., Z}{\expandafter\xdef\csname orcid\x\endcsname{\noexpand\href{https://orcid.org/\csname orcidauthor\x\endcsname} {\noexpand\orcidicon}} }

\begin{document}

\title{Approximate Invariant Analysis: An Efficient Framework for Nonlinear Beam Dynamics\\
{\small Part I: Geometric Approaches of the Poincar\'e Rotation Number}}

\author{Yongjun Li\orcidA{}}\thanks{email: yli@bnl.gov}
\affiliation{Brookhaven National Laboratory, Upton, New York 11973, USA}
\author{Sergei Nagaitsev\orcidB{}}
\affiliation{Brookhaven National Laboratory, Upton, New York 11973, USA}
\author{Derong Xu\orcidC{}}
\affiliation{Brookhaven National Laboratory, Upton, New York 11973, USA}
\author{Yue Hao\orcidD{}}\affiliation{Michigan State University, East Lansing, Michigan  48864, USA}
\author{Chad Mitchell\orcidE{}}
\affiliation{Lawrence Berkeley National Laboratory, Berkeley, California 94720, USA}

\begin{abstract}
We present the first part of an efficient framework for nonlinear beam dynamics, termed Approximate Invariant Analysis (AIA). The framework is based on the construction of approximate invariants~[Y.~Li, D.~Xu, and Y.~Hao, Phys.\ Rev.\ Accel.\ Beams \textbf{28}, 074001 (2025)] and on the extraction of the betatron frequency with the geometric foundations of Poincar\'e rotation number~[S.~Nagaitsev and T.~Zolkin, Phys.\ Rev.\ Accel.\ Beams \textbf{23}, 054001 (2020)]. The method is demonstrated using the National Synchrotron Light Source~II (NSLS-II) storage ring as an illustrative example.
\end{abstract}

\maketitle

\section{Introduction}
  In most existing ring-based particle accelerators, the presence of nonlinear magnets renders the machines non-integrable Hamiltonian systems. Consequently, exact analytical solutions for particle motion are generally unattainable because the system lacks a complete set of conserved quantities. A central goal of single-particle beam dynamics is therefore to assess particles' long-term stability.

  Traditional nonlinear beam-dynamics analysis is usually rooted in Hamiltonian perturbation theory. The linearized beam motion is first parameterized using Courant-Snyder theory~\cite{courant1958}. Once stable linear motion is established, nonlinear contributions are treated as perturbations. Common tools for studying nonlinear dynamics include, though are not limited to: (1) classical Hamiltonian perturbation theory, which employs canonical transformations to systematically average out fast oscillatory terms, thereby simplifying the Hamiltonian and highlighting the essential slow dynamics~\cite{wilson1995cas}; and (2) Lie-algebraic methods, which decompose nonlinear perturbations into resonance-driving terms~\cite{DragtLie}. Such perturbation methods can be extended to high order terms by combining with differential algebra~\cite{Berz1991a} and normal-form~\cite{chao2020lectures} techniques.

  Kolmogorov-Arnold-Moser (KAM) theory~\cite{Kolmogorov1954,Arnold1963,Moser1962} proves that long-term stable motion can persist in the presence of nonlinear perturbations, provided they remain sufficiently small. In such regions -- referred to as the dynamic aperture in accelerator physics -- quasi-periodic trajectories reside on invariant tori, which, although deformed, continue to inhibit chaotic diffusion. Reference~\cite{li2025construction} presents a method for constructing approximations to the invariant tori using Approximate Invariants (AI), enabling a qualitative assessment of particle stability by examining whether the resulting tori remain closed. Moreover, for a given torus, the associated quasi-constant betatron frequency can be extracted using the Poincar\'e rotation number (PRN) via a geometric approach~\cite{nagaitsev2020}. Building on these established foundations, this paper introduces an efficient framework for nonlinear beam dynamics analysis -- Approximate Invariant Analysis (AIA) -- and demonstrates its application using the National Synchrotron Light Source II (NSLS-II) storage ring~\cite{dierker2007}.

\section{Approximate invariant}
\begin{figure}
  \centering
  \includegraphics*[width=0.48\columnwidth]{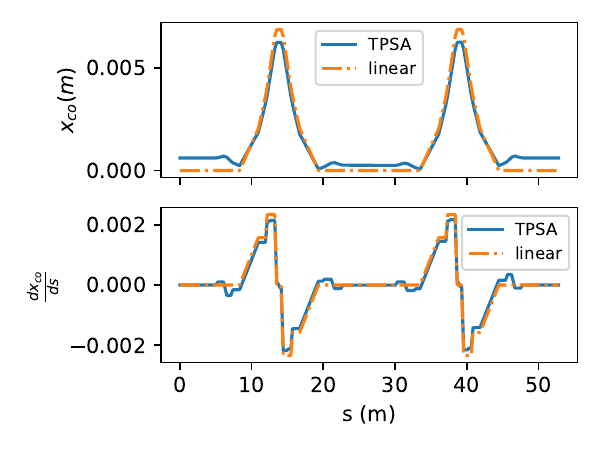}
  \includegraphics*[width=0.48\columnwidth]{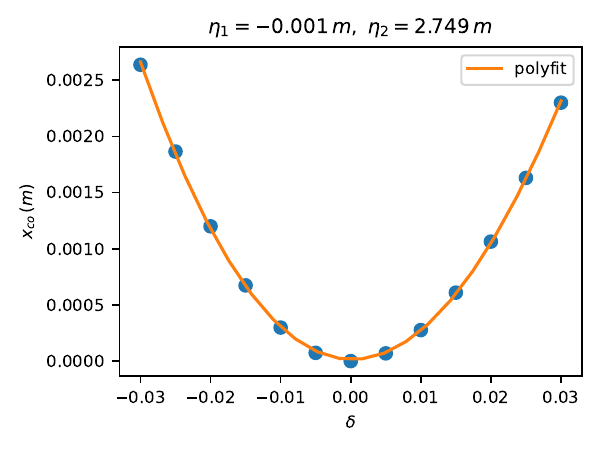} 
  \caption{Left: one NSLS-II super-cell's $\delta=1.5\%$ dispersive orbit (top) and its derivative (bottom). Solid blue lines include the nonlinear contribution from sextupoles and dash-dotted yellow lines represent the linear part $\mathbf{D}_1$. Right: different dispersive orbits observed at the longitudinal location $s=0$. The first order dispersion is near zero, because the NSLS-II lattice is a linear achromat.} 
  \label{fig:dispOrb}
  \end{figure}

  Although the concept of closed orbits is well known, we briefly revisit it for the sake of completeness of the overall framework. In accelerator physics, three types of closed orbits are commonly considered. An ideal transverse closed orbit coincides with the design trajectory, passing through the magnetic centers such that
  $\mathbf{X}^T = [x, p_x, y, p_y] = [0, 0, 0, 0]$ throughout the entire machine. The construction of approximate invariants (AIs) around the ideal reference orbit has been described in Ref.~\cite{li2025construction}. In the presence of magnetic imperfections and alignment errors, the actual closed orbit, denoted by $\mathbf{X}_{\mathrm{co}}$, deviates from the design trajectory. Furthermore, for an off-momentum particle, a momentum-dependent dispersive closed orbit is superimposed on this distorted orbit. Thus, a realistic closed orbit reads as
  \begin{equation}
      \mathbf{X}_{co}(\delta,s)\approx \mathbf{0}+\mathbf{X}_{co}(s)+\sum_{i=1,2,\cdots}\mathbf{D}_i(s)\delta^i,
  \end{equation}
  here $\mathbf{D}_i(s)$ is the $i^{th}$-order dispersion vector at $s$, $\delta=\frac{\mathrm{d}P}{P_0}$ is the relative momentum deviation\footnote{Strictly speaking, in the presence of an RF cavity system operating at a fixed frequency, variations in the closed-orbit path length lead to small changes in the beam momentum.}. Dispersive orbits can be obtained through iteration with re-normalized magnet bending or focusing strengths with $P = (1+\delta)P_0$. If there exist nonlinear magnets, it usually differs from the linear parameterization as shown in the left subplot of Fig.~\ref{fig:dispOrb}. High order dispersions can be obtained by a polynomial fitting as shown in the right subplot, or by implementing a differentiable map tracking.
  
\begin{figure}
  \centering
  \includegraphics*[width=0.8\columnwidth]{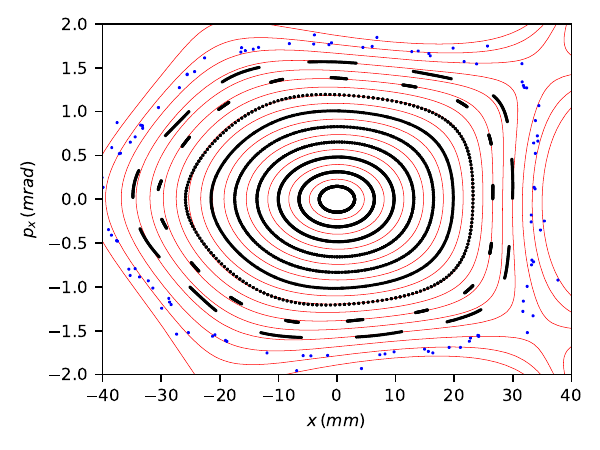}
  \caption{Contour plot of the horizontal AI of the NSLS-II ring in the mid-plane (solid red curves), overlaid with simulated Poincar\'e maps comprising 512 iterations. The simulations are shown as black dots for stable trajectories and blue dots for unbounded motion. Closed contours (tori) correspond to stable motion, whereas open contours indicate unbounded trajectories.} 
  \label{fig:tori}
  \end{figure}

  Once a closed orbit is established, one-turn maps in the form of truncated power series (TPS)~\cite{Berz1991a,yang2009array,zhang2024cpptpsa} can be obtained by tracking particles through the entire ring. Using the method described in Ref.~\cite{li2025construction}, two independent and Poisson-commuting AIs are constructed by computing the eigenvectors of the transpose of the one-turn transfer matrix expressed in square form. These AIs describe a family of (approximately) invariant tori in the transverse phase space $(x,p_x,y,p_y)$. In this paper, we only focus on the AI that predominantly characterizes the horizontal motion in the mid-plane, $y = p_y = 0$, as a representative example. For clarity, this constitutes a one Degree-of-Freedom (1-DoF) system in a two-dimensional phase space.
  \begin{equation}\label{eq:torus}
      \mathcal{K} = \sum_{mn}c_{m,n} x^m p_x^n.
  \end{equation}
  Equation~\eqref{eq:torus} describes a family of $x-p_x$ contours arising from the intersection of invariant tori with the mid-plane. The closure of these contours provides a qualitative measure of dynamical stability: closed curves (tori) correspond to stable motion, whereas open curves indicate unbounded or unstable trajectories. The outermost torus approximately delineates the boundary of the dynamic aperture. This qualitative stability criterion is further validated by tracking simulations, as illustrated in Fig.~\ref{fig:tori}.
  
\section{Betatron tune}
\begin{figure}
  \centering
  \includegraphics*[width=0.8\columnwidth]{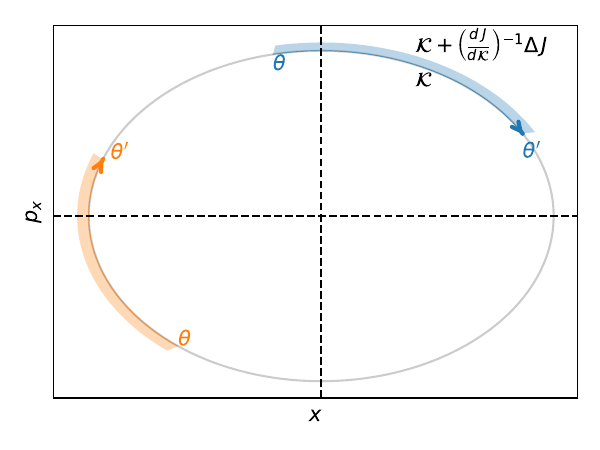}
  \caption{The derivatives of the partial action $J'$ with respect to the AI $\mathcal{K}$ under one-turn iteration are obtained by estimating the corresponding shadowed areas in phase space. Note that the one-turn map induces a clockwise rotation.} 
  \label{fig:partJ}
  \end{figure}

  AIs not only provide a qualitative criterion for dynamical stability, but also enable the extraction of the Betatron frequency using the geometric method proposed in Appendix~B of Ref.~\cite{nagaitsev2020} and its erratum~\cite{nagaitsev2026erratum}. In accelerator physics, this frequency -- defined as the number of oscillation cycles completed in one turn -- is referred to as the tune. Its fractional part often plays a critical role in determining proximity to resonant conditions. Here, we briefly outline the procedure for extracting the fractional tune using the concept of the rotation number, as defined by the following equation:
  \begin{equation}\label{eq:prn}
      \nu=\int_x^{x^\prime}{(\frac{\partial\mathcal{K}}{\partial p_x})^{-1}}\mathrm{d}x
      \bigg/
      \oint{(\frac{\partial\mathcal{K}}{\partial p_x})^{-1}}\mathrm{d}x
      =\frac{\mathrm{d}J^\prime}{\mathrm{d}\mathcal{K}} \bigg/ \frac{\mathrm{d}J}{\mathrm{d}\mathcal{K}},
  \end{equation}
  Here, the action integral~\cite{lichtenberg2013regular}
  \[
  J = \frac{1}{2\pi} \oint_{\mathcal{K}} p_x \, \mathrm{d}x
  \]
  is defined as a contour integral over the closed torus $\mathcal{K}$, while the partial action
  \[
  J' = \frac{1}{2\pi} \int_{x}^{x'} p_x \, \mathrm{d}x
  \]
  represents a sector integral corresponding to a single-turn iteration from $x$ to $x'$. Both integrals are evaluated along the AI tori defined by Eq.~\eqref{eq:torus}, as illustrated in Fig.~\ref{fig:partJ}.

  To evaluate the betatron tune using Eq.~\eqref{eq:prn}, we need to determine the constant-level sets of $\mathcal{K}$ defined by Eq.~\eqref{eq:torus}. In general, these level sets do not admit simple parameterization when $m+n \ge 4$. Nevertheless, they can be populated numerically with arbitrarily high density using a simple gradient-based minimization algorithm: starting from an initial guess of the form
  \[
  c_{2,0} x_0^{2} + c_{0,2} p_{x,0}^{2} = \mathcal{K},
  \]
  we compute the deviation from the target value of $\mathcal{K}$ and iteratively update the phase-space coordinates using local gradient information,
  \begin{equation}
  \nabla \mathcal{K}
  =
  \left[
  \sum_{m,n} m\,c_{m,n}\,x^{m-1} p_x^{n},
  \;
  \sum_{m,n} n\,c_{m,n}\,x^{m} p_x^{n-1}
  \right],
  \end{equation}
  until convergence is achieved.

  Once a sufficiently dense set of points on a torus has been populated, the derivative $\partial J / \partial \mathcal{K}$ can be evaluated conveniently in a polar coordinate system $(r,\theta)$. Specifically, we calculate the radial derivative of $\mathcal{K}$ with respect to $r$ at a fixed azimuthal angle $\theta$,
  \begin{align}
    \mathcal{K} &= \sum_{m,n} c_{m,n} x^m p_x^n
            = \sum_{m,n} c_{m,n} r^{m+n} \cos^m\theta \sin^n\theta, \nonumber\\
    \frac{\partial\mathcal{K}}{\partial r} &= 
    \sum_{m,n} (m+n)c_{m,n}\, r^{m+n-1} \cos^m\theta \sin^n\theta.
    \label{eq:rad_der}
  \end{align}
  Then, on the torus, the data points are ordered according to the azimuthal angle $\theta \in [-\pi,\pi]$. Between two neighboring points, the contribution to the action integral $J$ is approximated by
  \begin{equation}
    J_i \approx \frac{1}{2\pi}\frac{r_i^2(\theta_{i+1}-\theta_i)}{2}
         = \frac{1}{2\pi}\frac{r_i^2 \Delta\theta_i}{2}.
  \end{equation}
  Its derivative with respect to $\mathcal{K}$ is
  \begin{equation}
    \frac{\mathrm{d}J_i}{\mathrm{d}\mathcal{K}} = 
    \frac{r_i\Delta\theta_i}{2\pi}\frac{\mathrm{d}r_i}{\mathrm{d}\mathcal{K}}
    = \frac{r_i\Delta\theta_i}{2\pi}\left(\frac{\mathrm{d}\mathcal{K}}{\mathrm{d}r_i}\right)^{-1},
  \end{equation}
  when the gradient of $\mathcal{K}$ is predominantly radial,
  $\frac{\mathrm{d}\mathcal{K}}{\mathrm{d}r}\approx\frac{\partial\mathcal{K}}{\partial r}$, which has already been evaluated in Eq.~\eqref{eq:rad_der}. The denominator of Eq.~\eqref{eq:prn} becomes
  \begin{equation}
    \frac{\mathrm{d}J}{\mathrm{d}\mathcal{K}} = \sum_i \frac{\mathrm{d}J_i}{\mathrm{d}\mathcal{K}}.
  \end{equation}

  To calculate the numerator of Eq.~\eqref{eq:prn}, we choose a point $(x(r,\theta),\, p_x(r,\theta))$ on the torus and apply a one-turn iteration to obtain its image $(x'(r',\theta'),\, p_x'(r',\theta'))$. The derivative of the partial action is
  \[
  \frac{\mathrm{d}J'}{\mathrm{d}\mathcal{K}} = 
    \sum_{\theta' \le \theta_i \le \theta} \frac{\mathrm{d}J_i}{\mathrm{d}\mathcal{K}}.
  \]
  Because one-turn iterations advance clockwise, we always have $\theta' < \theta$, as annotated with the blue shaded area in Fig.~\ref{fig:partJ}. However, when $\theta'$ crosses the $-\pi$ boundary, the integration region becomes the union of two adjacent sectors, $[\theta, -\pi]$ and $[\pi, \theta']$, corresponding to the yellow shaded area.

\begin{figure}
      \centering
      \includegraphics*[width=0.8\columnwidth]{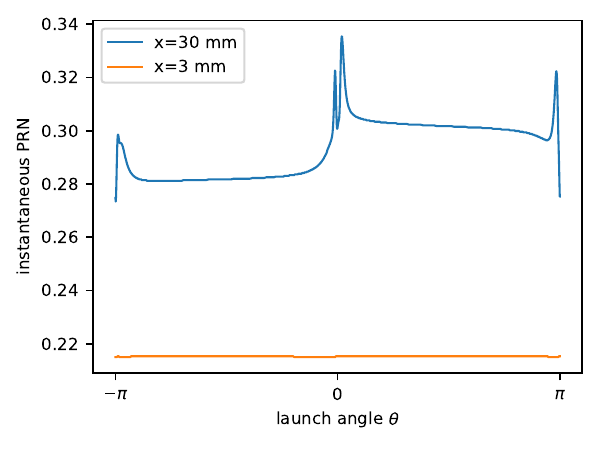}
      \caption{Variation of the instantaneous PRNs as a function of the initial azimuthal launch angle $\theta$ at two different amplitudes 3 and 30 \si{mm}. The periodicity of the betatron oscillation degrades with increasing amplitude.}  
      \label{fig:prnVariation}
  \end{figure}

  Since nonlinear Betatron oscillations are quasi-periodic, the instantaneous turn-to-turn rotation angle varies depending on the integration interval $[\theta,\theta']$. A quasi-constant tune is obtained by averaging over many successive iterations,
  \begin{equation}\label{eq:tune}
    \nu = \lim_{N\rightarrow\infty} \frac{1}{N} \sum_{i=1}^{N} \nu_i .
  \end{equation}
  Such a calculation formally requires successive iterations of the one-turn map, which can be computationally expensive. However, when the fractional tune is Diophantine irrational, the corresponding multi-turn trajectory becomes dense on the torus. In this case, the tune can be accurately evaluated by averaging a sufficiently large number of instantaneous PRNs, computed from initial conditions uniformly distributed along the AI tori. This tune calculation approach can be boosted by an efficient, embarrassingly parallel computation of PRNs as illustrated in Fig.~\ref{fig:prnVariation}.

\section{Amplitude-depended detuning}

\begin{figure}
  \centering
  \includegraphics*[width=0.9\columnwidth]{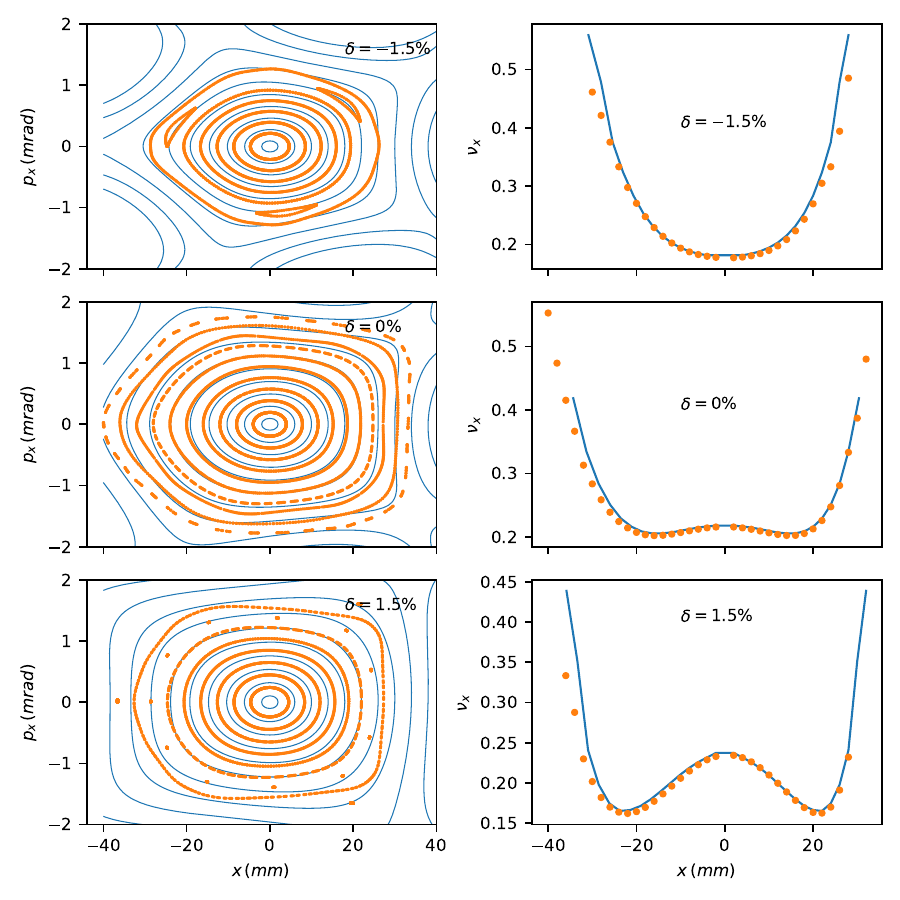}
  \caption{AI tori around different dispersive closed orbits (left column) and their ADDs (right column). The solid blue lines are calculated with AIA, while the yellow dots are obtained with tracking simulation.} 
  \label{fig:add}
  \end{figure}

  In nonlinear oscillators, amplitude-dependent detuning (ADD) can drive the tune to approach resonances, despite the linear tune being well separated from the resonance. In this section, we extract tunes directly on gradually increasing amplitude AI tori with the method described. Note that, unlike conventional approaches, in which detuning coefficients are typically estimated order-by-order, the proposed framework enables direct evaluation of the overall detuning associated with each torus. Figure~\ref{fig:add} presents the AIA-based estimates of the on- and off-momentum ($\delta = \pm 1.5\%$) dynamic apertures (left column) and the corresponding ADDs (right column) in the mid-plane of the NSLS-II ring. These estimates show reasonable agreement with results obtained from symplectic tracking simulations\cite{yoshida1990construction}.

\begin{figure}
  \centering
  \includegraphics*[width=0.8\columnwidth]{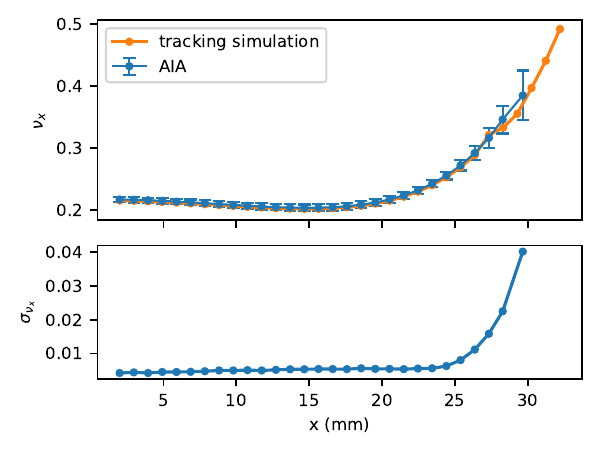}
  \caption{Variation of instantaneous PRNs at different amplitudes. The top subplot shows the average amplitude-dependent detuning, with PRN variations indicated by error bars. The bottom subplot highlights the sharp increase in PRN variations near the dynamic aperture boundary.} 
  \label{fig:tuneDiff}
  \end{figure}

  As mentioned previously, even when the motion is confined to a single torus, instantaneous PRNs vary with azimuthal angles as illustrated in Fig.~\ref{fig:prnVariation}. When PRNs are averaged over successive $N$-turn intervals, the differences between these averages quantify the sensitivity of the tune to initial conditions, commonly referred to as tune diffusion~\cite{laskar2003,papaphilippou2014}. However, this procedure requires repeated iterations of the one-turn map and is therefore not computationally more efficient than the Numerical Analysis of Fundamental Frequencies (NAFF) method. In contrast, we observe that PRNs evaluated from non-successive iterations with randomly chosen initial conditions exhibit pronounced variation as particle motion approaches the boundary of the dynamic aperture, as highlighted in the bottom panel of Fig.~\ref{fig:tuneDiff}. These variations provide a clear signature of the breakdown of quasi-periodicity and, consequently, the loss of long-term stability. Similar behavior has previously been exploited as a fast chaos indicator~\cite{szezech2013finite}, and it may offer a promising tool for nonlinear lattice optimization.
  
\section{Conclusion}

  We have demonstrated an efficient nonlinear beam-dynamics framework based on approximate invariant reconstruction and betatron frequency extraction, using the NSLS-II storage ring as a representative example. Within this framework, the analysis does not rely on conventional Hamiltonian perturbation theory, Lie-algebraic methods, or normal-form techniques, nor does it require Courant-Snyder linear parameterization. Instead, key dynamical quantities -- such as the dynamic aperture and amplitude-dependent detuning -- can be evaluated directly and efficiently, with results that are validated by symplectic tracking simulations. In this paper, this analysis approach is applied to a 1-DoF system, specifically the horizontal motion in the mid-plane. Nevertheless, it is readily extensible to N-DoF systems, for example by adopting the formulation described in Ref.~\cite{mitchell2021}. An alternative method for extracting betatron tunes in N-DoF systems, based on time-of-flight analysis for approximate invariant flows~\cite{xu2025frequency}, will be presented in a forthcoming companion paper (Part~II).
  
\section*{Data availability}
  The data that support the findings of this study  are available upon reasonable request and subject to standard U.S. national laboratory data-sharing policies.

\begin{acknowledgments}
  We would like to thank, Y-K.~Kan, T.~Shaftan and V.~Smaluk (BNL) for the discussion and support. This research is supported by the U.S. Department of Energy (DOE) under Contract No. DE-SC0012704, and the DOE Basic Energy Sciences (BES) Field Work Proposal (FWP) 2025-BNL-PS040, the DOE's Early Career Program, and the DOE High Energy Physics (HEP) award DE-SC0019403.
\end{acknowledgments}

\bibliography{ref.bib}

\end{document}